# The Report on China-Spain Joint Clinical Testing for Rapid COVID-19 Risk Screening by Eye-region Manifestations

Yanwei Fu[1*], Feng Li[2*], Paula boned Fustel[6*], Lei Zhao[7*], Lijie Jia[12*], Haojie Zheng[9*], Qiang Sun[4*], Shisong Rong[8*], Haicheng Tang[2*], Xiangyang Xue[4*], Li Yang[9*], Hong Li[10*], Jiao Xie[7*] Wenxuan Wang[3], Yuan Li[9], Wei Wang[9], Yantao Pei[9], Jianmin Wang[9], Xiuqi Wu[9], Yanhua Zheng[9], Hongxia Tian[9], Mengwei Gu[1,5,#]

Affiliations

1. School of Data Science, Fudan University, Shanghai, 200433, China
2. Shanghai Public Health Clinical Center, Fudan University, Shanghai, 201508, China
3. School of Computer Science, Fudan University, Shanghai, 200433, China
4. Academy for Engineering & Technology, Fudan University, Shanghai, 200433, China
5. Aimomics (Shanghai) Intelligent Technology Co., Ltd., Shanghai, 200433, China
6. University and Polytechnic Hospital La Fe, Valencia, 46026, Spain
7. Department of Infectious Diseases, Union Hospital, Tongji Medical College, Huazhong University of Science and Technology, Wuhan, 430022. China.
8. Department of Ophthalmology, Harvard Medical School, Massachusetts Eye and Ear, Mass General Brigham, Boston, USA
9. The Fifth Hospital of Shijiazhuang, Hebei Medical University, Shijiazhuang, China
10. Medical Examination Center, Hubei Provincial Hospital of Traditional Chinese Medicine, Wuhan, 430061, China
11. Department of respirology, Shanghai Public Health Clinical Center, Fudan University, Shanghai, 201508, China
12. Department of Anesthesia, the International Peace Maternity and Child Health Hospital, Shanghai Jiao Tong University School of Medicine, 200025, Shanghai, China

* These authors contribute equally to this study.
# Corresponding author: Mengwei Gu

Short title: Eye-image-based AI rapid screening for COVID-19

Abstract

Background

The worldwide surge in coronavirus cases has led to the COVID-19 testing demand surge. Rapid, accurate, and cost-effective COVID-19 screening tests working at a population level are in imperative demand globally.

Methods

Based on the eye symptoms of COVID-19, we developed and tested a COVID-19 rapid prescreening model using the eye-region images captured in China and Spain with cellphone cameras. The convolutional neural networks (CNNs)-based model was trained on these eye images to complete binary classification task of identifying the COVID-19 cases. The performance was measured using area under receiver-operating-characteristic curve (AUC), sensitivity, specificity, accuracy, and F1. The application programming interface was open access.

Findings

The multicenter study included 2436 pictures corresponding to 657 subjects (155 COVID-19 infection, 23·6%) in development dataset (train and validation) and 2138 pictures corresponding to 478 subjects (64 COVID-19 infections, 13·4%) in test dataset. The image-level performance of COVID-19 prescreening model in the China-Spain multicenter study achieved an AUC of 0·913 (95% CI, 0·898-0·927), with a sensitivity of 0·695 (95% CI, 0·643-0·748), a specificity of 0·904 (95% CI, 0·891 -0·919), an accuracy of 0·875(0·861-0·889), and a F1 of 0·611(0·568-0·655).

Interpretation

The CNN-based model for COVID-19 rapid prescreening has reliable specificity and sensitivity. This system provides a low-cost, fully self-performed, non-invasive, real-time feedback solution for continuous surveillance and large-scale rapid prescreening for COVID-19.

Funding

This project is supported by Aimomics (Shanghai) Intelligent

Introduction

The Coronavirus disease 2019 (COVID-19) has affected more than a billion people with unprecedented public health and economic costs. In the battle of controlling COVID-19 infections, the first and critical step is to identify new outbreaks of coronavirus infections. Using polymerase chain reaction (PCR) based nucleic acid detection is the "golden standard" with high sensitivity and good specificity. However, the PCR detection requires significant time and hardware investments which can be prohibitory for rapid screening purposes. Therefore, several rapid screening tests have been utilized to identify potential COVID-19 cases, including direct body temperature measurement, symptom assessment, travel and exposure history review, and combinations of the above-mentioned tests. However, all these rapid screening tests suffer from a hugely variable sensitivity in detecting COVID-19 cases[1]. Approximately 80% of the published studies reported a sensitivity of <50% in different target populations and scenarios[2]. Therefore, a rapid, accurate, and cost-effective COVID-19 screening tests working at a population level is in imperative demand globally.

    A spectrum of extrapulmonary manifestations of COVID-19 have been reported, including ocular symptoms[3,4]. These ocular symptoms consist of a wide range of manifestations in the eyelid, ocular surface, anterior segment, and the posterior segment of the eye[5-8]. The prevalence of ocular symptoms in COVID-19 cases reportedly ranged from 0 to 38% with an average prevalence of 11%[9]. The ocular manifestations as first presenting symptom or only symptom of COVID-19 were reported in up to 10% of COVID-19 cases[5,6,10]. These findings suggested that ocular symptoms could manifest specific features in symptomatic and even in presymptomatic COVID-19 patients. Although the clinical significance of ocular symptoms has yet to be fully elucidated[9-11], the typical ocular symptoms provide useful information to recognize the potential infectious cases.

    Artificial intelligence (AI) and machine learning, particularly image-based deep-learning models with convolutional neural networks (CNNs), have shown promising performance in disease classifications and risk assessments[12,13]. However, these AI systems require Computed Tomography scans in hospital settings, which are not available for screening tests at a population level. In this study, we developed and tested a rapid eye-image-based AI system, setting in a smartphone, to identify potential COVID-19 infection in three independent cohorts including two from China and one from Spain.

Methods

Ethical statements

The study was conducted in accordance with of the Declaration of Helsinki. The study protocol was approved by the Ethics Committee of Shanghai Public Health Clinical Center, Fudan University (approval No.: YJ-2020-S078-02), the fifth hospital of Shijiazhuang affiliated to Hebei Medical University (approval No.: 2021005), and the Ethics Committee on Drug Research of the University and Polytechnic Hospital La Fe (registry No: 2020-637-1). Informed consents were obtained from all participating individuals.

Study design and participants

This multicentral study was led by the Shanghai Public Health Clinical Center (SPHCC), Fudan University, Shanghai, China. The study was conducted in two phases to enroll participants from three hospitals: the Department of Respiratory and Critical Care Medicine (DRCCM) of SPHCC, the fever isolation ward of the Fifth Hospital of Shijiazhuang (FHS), and the Department of Ophthalmology at the University and Polytechnic Hospital La Fe (DOUPH). The SPHCC is a tertiary class A general hospital, a WHO and national clinical research and training center for emerging and reappearing infectious diseases. It is the government-designated hospital for the management of COVID-19 cases in Shanghai. The FHS is the first tertiary hospital for treatment, research, and prevention of infectious diseases in Hebei province, China. It is the national demonstrative base for prevention and treatment of liver diseases, the quality control center in Hebei province for diagnosis and treatment of AIDS, and the designated hospital for diagnosis and treatment of COVID-19 cases in Shijiazhuang city, China. Fever isolation ward is an emergency department established during the pandemic. The Department of Ophthalmology at the University and Polytechnic Hospital La Fe is a leading eye care provider, educator and researcher in Spain. All the COVID-19 patients were enrolled by DOUPH through collaboration with department of microbiology.

The first phase of data collection was completed from April $1^{st}$ 2020 to June $7^{th}$ 2021 at the SPHCC and DOUPH, which generated the development dataset. Development dataset was randomly divided into an independent training dataset and a validation dataset after considering the sample sizes and COVID-19 cases in each dataset. The training dataset was used to develop the model and the validation dataset was used to determine the best hyper-parameters and threshold (a score above the threshold is classified as positive). The second phase of the data collection was completed during June $1^{st}$ 2020 and June $11^{th}$ at the SPHCC, FHS and DOUPH. This dataset was used to test the performance of the models.

Controls were enrolled from each center during the corresponding phases of enrollment. The control dataset from SPHCC included subjects with non-COVID-19 lung diseases (NCLD), ocular diseases and healthy volunteers. NCLD included bronchopneumonia, chronic obstructive pulmonary diseases, pulmonary fungal infection and lung cancer. Ocular diseases included trachoma, pinkeye, conjunctivitis, glaucoma, cataract and keratitis. All healthy volunteers received physical examinations, had no abnormal clinical findings, and had no contact history with COVID-19 patients. In the FHS dataset, controls included patients fully recovered from COVID-19 (nuclear acid negative), patients with HBV (HBV DNA positive) and healthy subjects. In DOUPH dataset. All the control subjects were tested negative for COVID-19 nucleic acids during the study period. No death events were observed in this study.

Image acquisition and preprocessing

We conducted a study on the CNN-based model, using eye-region photos to screen for COVID-19 patients (*Figure 1*). For each participant, normally 3-5 photographs of the ocular surface were taken using smart mobile phone CCD and CMOS cameras, assisted by doctors or healthcare workers. The same shooting mode and parameters were used during image capturing. And all shooting filters were avoided. All photos were captured in a good lighting condition, and not in a dark or red background. The image resolution was at least 1900x500 pixels at 96dpi. The average time for taking a set of 5 eye photos was around 1 minute. In the re-examination step, all the photos are examined by human for the second time, the data that failed to reveal the details of the eyes were discarded.

In this study, the images from Shanghai were captured by different mobile photos, including HONOR Play 3, MAIMANG 8, HUAWEI Mate 9, OPPO R15, HONOR Magic 2s, and iPhone 8Plus. The images from Hebei were captured by Xiaomi 8, iPhone 7Plus, Huawei PRA-AL00X and Huawei ANA-AN00. The images from Spain were captured by OPPO A37f, Samsung SM-A920F and iPhone SE $2^{nd}$.

Development of the classification network

The schematic of our proposed model is illustrated in *Figure 2*, which consists of two components, the Image Preprocessing[14] and Classification Networks. Specifically, the Image Preprocessing received raw eye-region images, and prepared them for model training or inference. The Classification Network was built upon the deep learning architecture for the classification. It studied the characteristics of eye-region according to the inputs, and learnt discriminatory texture and semantic embeddings in a high-dimension space. Finally, the risk assessment of COVID-19 was predicted. During training, to avoid overfitting, the random crop and early stop strategies were applied.

The developing process of classification network had been divided into training and testing stages as shown in Fig. 1. In the training stage, the classification network was evaluated on the validation dataset and was used to determine the hyperparameters, the threshold was determined by the best F1 score. In the testing stage, the model's performance (AUC, sensitivity, specificity, accuracy, F1) was measured using the test dataset with the determined threshold. Considering that a subject may have more than one available image, we classified each subject based on the prediction results of multiple images[15]. Therefore, we conduct the risk screening for subject based on the previous image-level predictions. To enhance the robustness or sensitivity, we developed two vote strategies: mean-voting and max-voting were applied. In the mean-voting method, the image-level prediction scores were averaged to the final score. In the max-voting method the highest image-level prediction score was used as the final score which resulted in a higher sensitivity.

Statistical analysis

To measure the performance of the binary classification network, we calculated the area under the receiver-operating-characteristic curve (AUC), sensitivity, specificity, and accuracy according of our classification network. Bootstrapping with 1000 replicates was used to estimate 95% confidence intervals of the performance metrics, with the photo as the resampling unit. In addition, the receiver operating characteristic curves (ROCs) was plotted to illustrate the performance in screening COVID-19 disease.

Results

Study subjects

In the cohort from Shanghai, the participants were enrolled at the Shanghai Public Health Clinical Center (SPHCC), Fudan University and AIMOMICS. In the development datasets, 104 COVID-19 patients, 342 control group participants (143 healthy volunteers, 131 NCLD, 68 OD patients) were recruited during 2020 April $1^{st}$ to June $30^{th}$. The test dataset comprised of 29 COVID-19 patients and 99 control group participants (35 healthy volunteers, 31 NCLD, 33 OD patients) who were enrolled during 2020 July $1^{st}$ to August $31^{st}$ (Table 1). Among the 133 COVID-19 patients, 47 (including 24 in the development and 23 in the test dataset) were asymptomatic/mild cases. And the majority of the participants were East Asian (87·50% of the development dataset and 93·10% of the test dataset).

In the Hebei cohort, the participants were enrolled from the Fifth Hospital of Shijiazhuang, Hubei Provincial Hospital of Traditional Chinese Medicine and AIMOMICS. In the development datasets, 20 COVID-19 patients were enrolled during 2021 Jan $1^{st}$ to Jan $23^{th}$. The test dataset was comprised of 27 COVID-19 patients and 161 controls who were enrolled during 2021 Feb $1^{st}$ to April $30^{st}$ (Table 1).

In the Spain cohort, the participants were enrolled from the La Fe University, Polytechnic Hospital and AIMOMICS. In the development datasets, 31 COVID-19 patients and 160 controls were enrolled during 2020 November $1^{st}$ to 2021 June $7^{th}$. The test dataset comprised 8 COVID-19 patients and 154 controls who were enrolled during 2021 January $1^{st}$ to June $11^{st}$ (Table 1).

The demographic characteristics of COVID-19 patients were shown in Table 1. The Shanghai cohort had 2,108 photographs (development 1561, test 547) of 574 participants (development 446, test 128). The data in Hebei included 1041 photographs (development 101, test 940) of 208 participants (development 20, test 188). The data in Spain included 1426 photographs

(development 774, test 652) of 353 participants (development 191, test 162). In total, there were 902 photographs of 219 COVID-19 cases.

Results on the test dataset of China and Spain

The classification network was trained on the multicenter training dataset. The classification network achieved an AUC of 0·953 (95%CI, 0·936-0·969) in the Shanghai test dataset, an AUC of 0·866 (95%CI, 0·837-0·895) in the Hebei test dataset, and an AUC of 0·925 (95%CI, 0·873-0·976) in the Spain test dataset at image-level. The AUC of both max-voting and mean-voting strategies were shown in Table 2. The ROCs of image-level, max-voting, and mean-voting were plotted in Figure 3.

In the multicenter study, the 302 photos of COVID-19 patients have 210 (69·5%) classified correctly, the 1841 negative photos have 1665 (90·5%) classified correctly. With max-voting strategy, we can correctly classify 61 COVID-19 patients from 64, while the mean-voting strategy successfully detected 50 COVID-19 patients. The confusion matrix of classification results of subject-level and image-level were shown in Table 3. However, misjudgments were also documented, The false negative rate is 30·5%, 21·9%, and 4·7% with respect to the image-level, mean-voting level and max-voting level.

Discussion

Typical ocular symptoms of COVID-19 could be captured by our specific AI model. The model successfully differentiated COVID-19 patients from non-COVID-19 controls, with higher specificity and sensitivity. Notably, the inputted eye-region images were captured by general cellphone cameras, which underlined the excellent convenience and easy accessibility of the ocular photos-based prescreening system for clinical translation.

A meta-analysis demonstrated that the four most common ocular symptoms/signs were follicular conjunctivitis, redness, watering, and discharge[16]. Conjunctivitis could be the sole symptom of COVID-19[17]. Although there is a low prevalence of SARS-CoV-2 in tears collected from conjunctival swabs, it is possible to transmit via the eyes[18]. Therefore, screening of patients with conjunctival congestion by ophthalmologists was advocated during the outbreak of COVID-19[19]. In this study, we provide a non-contact method to screening the eye's manifestations, not only for ophthalmologists, but also for ordinary people. Besides the conjunctival manifestations, the SARS-CoV-2 can affect the inner and outer retinal layers, which might also contribute to the ocular symptoms, such as inflammation and ocular pain[20]. Knowledge of eye symptoms and ocular transmission of the virus remains incomplete. However, based on the clinical findings, the implementation of innovative changes such as AI may assist in battling against the COVID-19 infection[13].

The current model has two major components: an image preprocessing method to detect, crop, and align the eye area from the input image, and a CNN-based screen model to extract discriminative features and recognize COVID-19 patients based on the processed eye-region data. The extracted high-dimensional feature is used to compare with the prototype features in the knowledge base for classification. Because of the great ability to capture the specific eye features, the methods have high specificity when the control and the positive are different in the eye area (for example, COVID-19 patients might have ocular symptoms). On the other hand, some COVID-19 patients might have no obvious ocular symptoms in some viewpoints, this makes the model might have a low sensitivity on a single image. Thus, the voting strategies on multiple images are adopted in our paper to boost the robustness of the model.

As shown in Figure 4, we generate the average heatmaps on the 5 typical poses (forward, left, up, right, and down respectively) to have an overall understanding about the attention of our model. Specifically, we randomly picked 20 images on each pose and group. The average heatmaps show that the eye features have obvious differences and certain regularity. In order to increase the interpretability of the model, we visually analyzed the key areas of the model's attention in the classification process. Concretely, the key areas of the model's attention were converted into heat maps based on gradients and activation maps by GradCAM. Specifically, GradCAM has been successfully applied to fast detection of COVID-19 cases by chest X-ray and CT-Scan images[1], which helps the human to better understand the predictions of the deep learning model. (1) The classification network extracted features to Non-COVID-19 participants have evenly distributed attention on the eye area; (2) The ocular features to patients with COVID-19 mainly focus on the inner and outer corner of the eye, in addition it covers upper, lower eyelid and other eye area. These features might suggest the specific site in eyes where inflammation or immune reaction happened in COVID-19 patients, which need further investigations.

For case study heatmap in *Figure 5*, we found that the attention of heatmaps of Non-COVID-19 group are around the sclera and eyelid in the row 1 and row 2. The attention heatmap of COVID covers upper and lower eyelid, the inner eye corner and especially the outer eye corner in row 1 and row 2.

To test whether the model can distinguish the asymptomatic COVID-19 patients from the control group. We leave only the 23 asymptomatic COVID-19 patients in the shanghai dataset (6 non-asymptomatic COVID-19 patients are removed) and make an asymptomatic COVID-19 vs. Non-COVID-19 test. The sensitivity is 72·8%, 100%, 87% on image-level, max-voting subject-level and mean-voting subject-level, which is almost the same as the sensitivity of the default COVID-19 vs Non-COVID-19 test (71·2%, 100% and 86·2% on image-level, max-voting subject-level and mean-voting subject-level). This implies that the model can effectively distinguish the asymptomatic COVID-19 patients from the control group.

In addition, this study belongs to a long-term global project. The project would gradually open more than 300 diseases eye tests, such as virus influenza, diabetes, hepatopathy, etc. through the open accessible APIs. The test algorithm could be easily deployed or embedded in the high-definition (HD) camera and any detect accessories, combined into a multi-modal approach including vision and other sensors, continually monitoring the particular disease control areas such as the transportation hub, population center, quarantine house, making the health care more accessible with lower cost. We believe this system could be easily realized with HD qualified eye-region images and selfies for rapid COVID-19 prescreening. Moreover, the current study could be inspiring and helpful for encouraging more researches on this topic.

There are some limitations in this study. First, the participants were mostly collected from East Asia (China) and some from Spain. Therefore, a larger multicenter study covering more patients with diverse races and more control groups is necessary before the model could be used globally. More data are being collected and will be used in the further study. Secondary, there might be potential confounding factors such as comorbidities influencing eye symptoms. We did not collect the comorbidities of our participants. However, the control and positive patients were randomly selected from the population, which could balance the baseline demographics between groups. Third, some of the demographic information (e.g., gender and ages) are not collected during image acquisition. Fourth, our model was based on the eye symptoms, however, it cannot determine COVID-19-related eye disease. The pathological significance of extracted features from COVID-19 patients should be carefully interpreted and re-verified by the ophthalmologist. Further clinical studies are needed to test the performance and provide a deeper understanding of our findings of the ocular surface feature-based classification network.

CONCLUSION

In this study, the rapid COVID-19 screening model with the CNN based on eye-region images captured by typical cellphones or qualified selfies had high specificity and acceptable sensitivity. As an available rapid solution of fully self-performed prescreening in turnaround time, capabilities include the lower cost, fully self-performed, non-invasive, real-time results, continuous surveillance, and open accessible APIs. Anyone anywhere anytime can use cellphone eye self-portraits to tell the risk probability and get the result within 1 minute. We believe a system implementing such an algorithm should assist the large-scale rapid screening for COVID-19 infection.

Declarations

public health clinic center of Fudan University. which conducted in Spain was approved by the Ethics Committee on Drug Research of the University and Polytechnic Hospital La Fe.

Data & Code Availability

Not publicly available due to NDA & AIMOMICS license protection.

# Figures

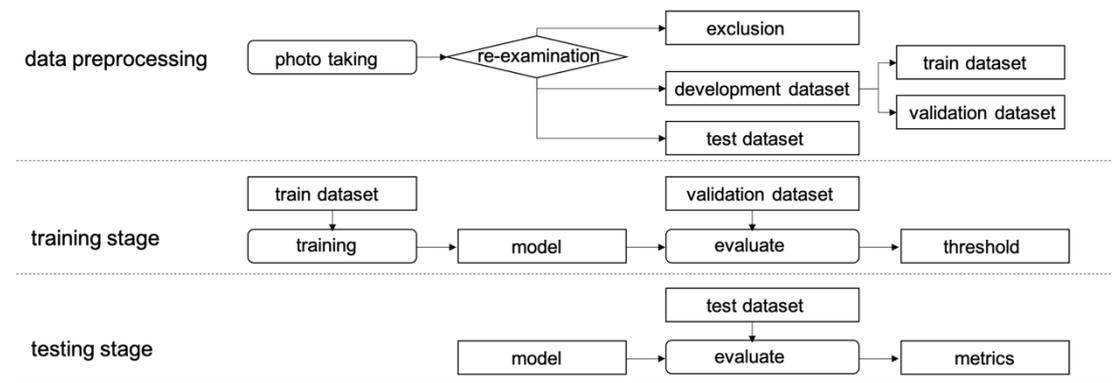

**Figure 1. Study design and workflow of this study.**

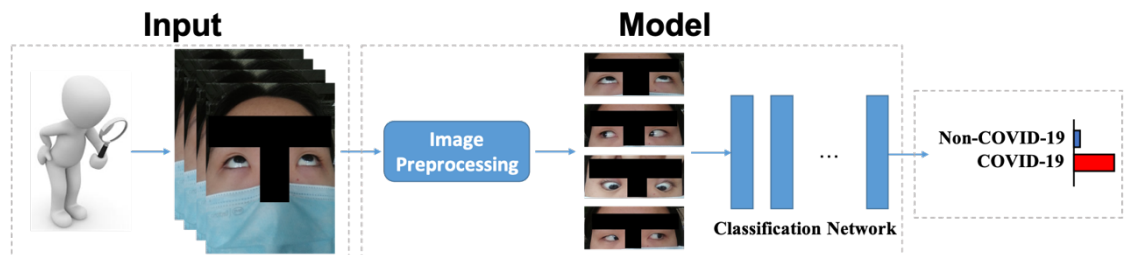

**Figure 2. Illustration of the framework.**

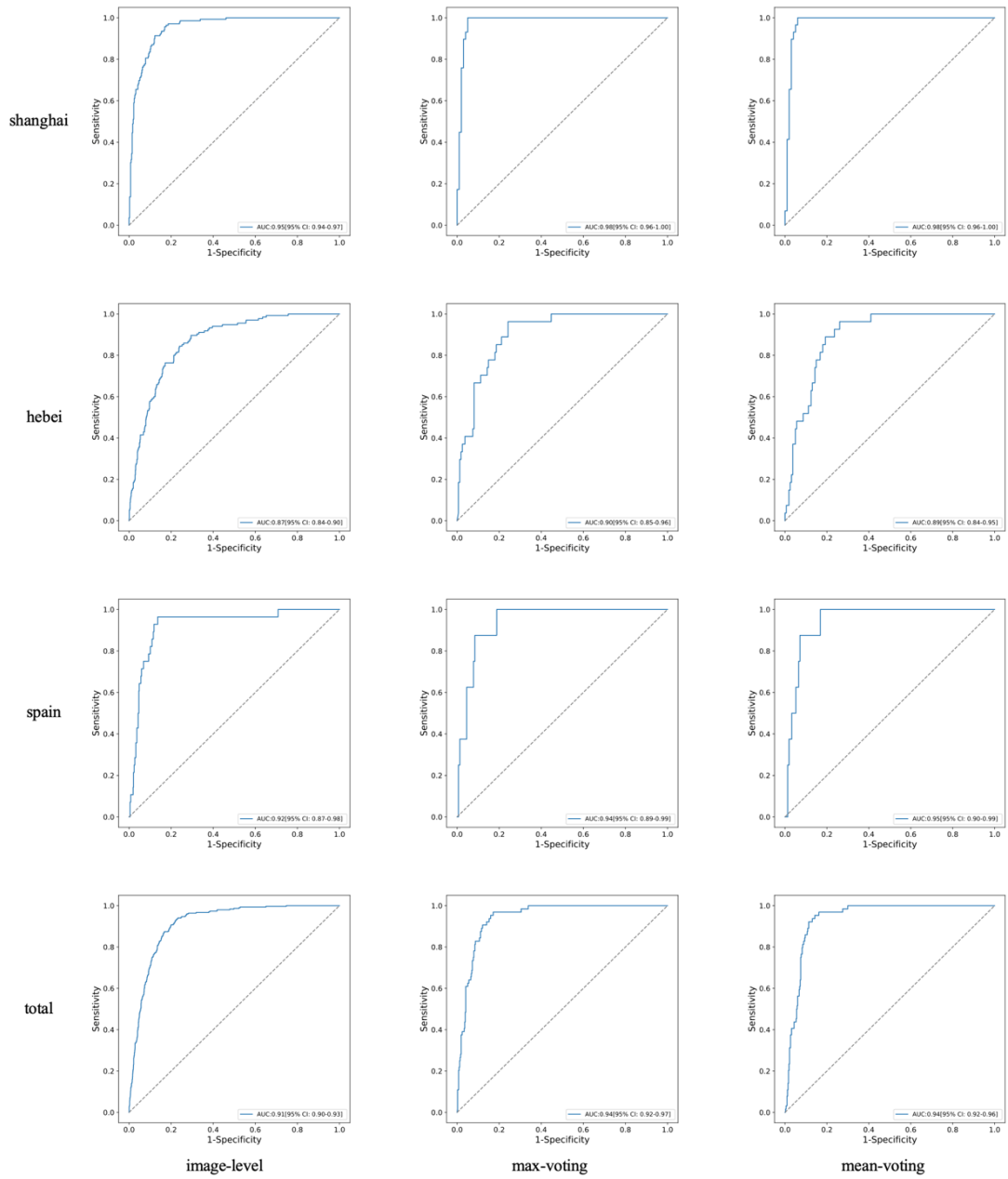

Figure 3. COVID-19 vs. Non-COVID-19 ROC curves on the test datasets of Shanghai, Hebei, Spain and the Total.

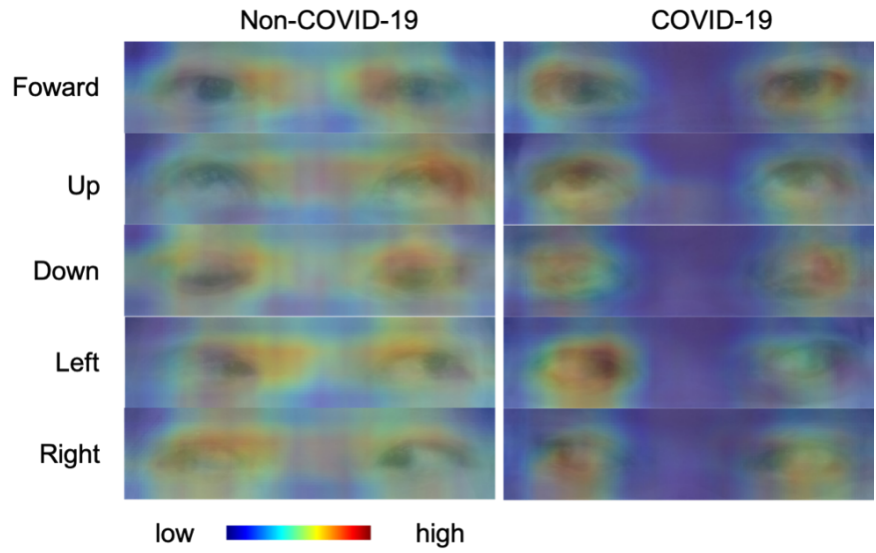

Figure 4. Average heatmaps.

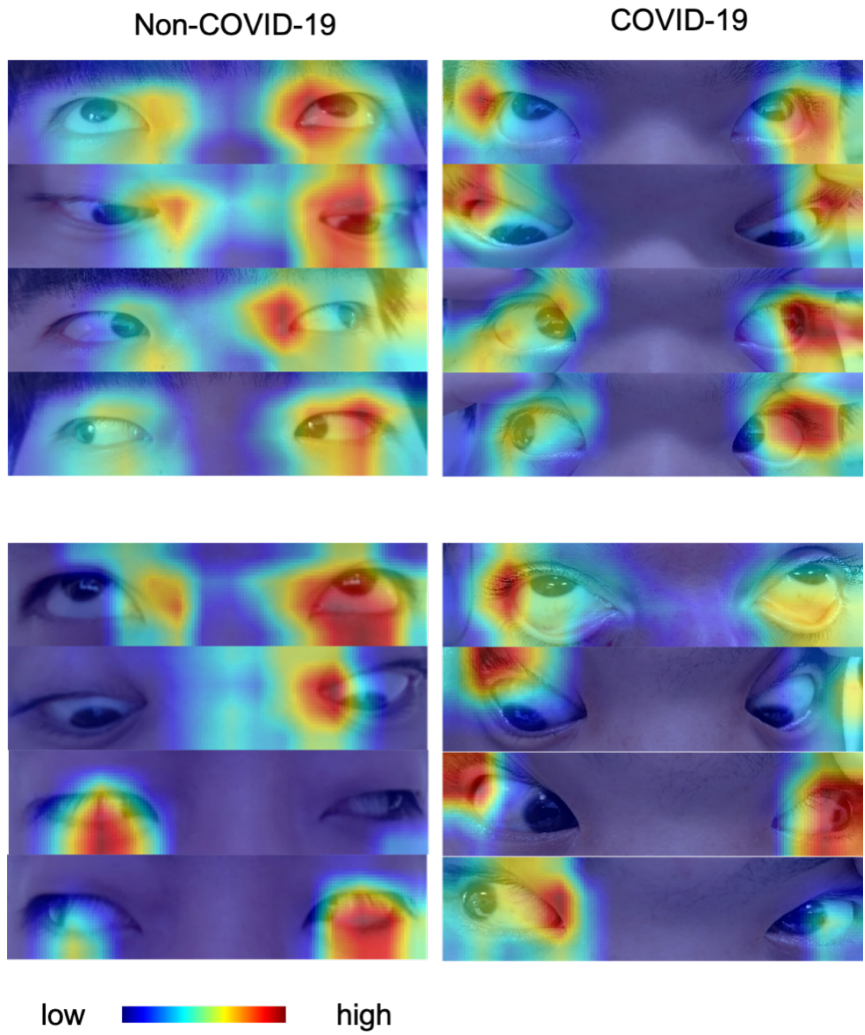

Figure 5. Case heatmaps.

## Tables

### Table 1. Summary of Development(Training/Validation), and Testing Datasets

|  |  | Shanghai | | Hebei | | Spain | | Total | |
|---|---|---|---|---|---|---|---|---|---|
|  | Infor. | Develop. | Test | Develop. | Test | Develop. | Test | Develop. | Test |
| Data aquisition | date | 2020·4·1 - 2020·6·30 | 2020·6·1- 2020·8·31 | 2021·1·1- 2021·1·23 | 2021·2·1- 2021·4·30 | 2020·11·1- 2021·6·7 | 2021·1·1- 2021·6·11 | 2020·4·1 - 2021·6·7 | 2020·6·1- 2021·6·11 |
| COVID-19 | Subject | 104(83/21) | 29 | 20(14/6) | 27 | 31(22/9) | 8 | 155(119/36) | 64 |
|  | Images | 367(290/77) | 139 | 101(68/33) | 135 | 132(96/36) | 28 | 600(454/146) | 302 |
| Age | Years | 5-65 | 20-59 | 20-55 | 20-55 | 21-94 | 25-69 | 5-94 | 20-69 |
| Sex | Male | 71(68·27%) | 21(72·41%) | 4(20%) | 11(40·7%) | 17(54·8%) | 1(12·5%) | 87 | 22 |
|  | Female | 33(31·73%) | 8(27·59%) | 16(80%) | 16(59·3%) | 14(45·2%) | 7(87·5%) | 47 | 15 |
| Control Group | Subject | 342(272/70) | 99 | 0 | 161 | 160(119/41) | 154 | 502(391/111) | 414 |
|  | Images | 1194(958/236) | 408 | 0 | 805 | 642(482/160) | 624 | 1836(1440/396) | 1837 |
| Total | subject | 446(355/91) | 128 | 20(14/6) | 188 | 191(141/50) | 162 | 657(510/147) | 478 |
|  | images | 1561(1248/313) | 547 | 101(68/33) | 940 | 774(578/196) | 652 | 2436(1894/542) | 2139 |

Infor.=information. Develop.=development.

### Table 2. Classification Performance of the classification network on the test dataset of Shanghai, Hebei, Spain and Total

|  | AUC(95% CI) | Sensitivity(95% CI) | Specificity(95% CI) | ACC(95% CI) | F1(95% CI) |
|---|---|---|---|---|---|
| **Image-Level (Shanghai)** | | | | | |
| COVID-19 vs. Non-COVID-19 | 0·953(0·936-0·969) | 0·712(0·634-0.791) | 0·946(0·924-0·968) | 0·887(0·860-0·913) | 0·762(0·701-0·822) |
| **Subject-Level (Shanghai)** | | | | | |
| COVID-19 vs. Non-COVID-19 (Max-Voting) | 0·982(0·961-1·000) | 1·000(1·000-1·000) | 0·869(0·799-0·938) | 0·898(0·845-0·952) | 0·817(0·716-0·918) |
| COVID-19 vs. Non-COVID-19 (Mean-Voting) | 0·979(0·955-1·000) | 0·862(0·732-0·992) | 0·970(0·935-1·000) | 0·945(0·905-0·986) | 0·877(0·782-0·972) |
| **Image-Level (Hebei)** | | | | | |
| COVID-19 vs. Non-COVID-19 | 0·866(0·837-0·895) | 0·741(0·668-0·814) | 0·839(0·812-0·865) | 0·824(0·800-0·849) | 0·548(0·485-0·611) |
| **Subject-Level (Hebei)** | | | | | |
| COVID-19 vs. Non-COVID-19 (Max-Voting) | 0·903(0·851-0·955) | 0·963(0·885-1·000) | 0·733(0·662-0·803) | 0·766(0·705-0·827) | 0·542(0·420-0·664) |
| COVID-19 vs. Non-COVID-19 (Mean-Voting) | 0·894(0·842-0·946) | 0·778(0·611-0·945) | 0·845(0·787-0·902) | 0·835(0·779-0·891) | 0·575(0·434-0·717) |
| **Image-Level (Spain)** | | | | | |
| COVID-19 vs. Non-COVID-19 | 0·925(0·873-0·976) | 0·393(0·209-0·577) | 0·963(0·948-0·979) | 0·938(0·919-0·958) | 0·355(0·199-0·511) |
| **Subject-Level (Spain)** | | | | | |
| COVID-19 vs. Non-COVID-19 (Max-Voting) | 0·942(0·890-0·993) | 0·750(0·420-1·000) | 0·922(0·879-0·965) | 0·914(0·869-0·958) | 0·462(0·216-0·707) |
| COVID-19 vs. Non-COVID-19 (Mean-Voting) | 0·946(0·899-0·992) | 0·500(0·119-0·881) | 0·961(0·929-0·993) | 0·938(0·901-0·976) | 0·444(0·149-0·739) |
| **Image-Level (Total)** | | | | | |
| COVID-19 vs. Non-COVID-19 | 0·913(0·898-0·927) | 0·695(0·643-0·748) | 0·904(0·891-0·919) | 0·875(0·861-0·889) | 0·611(0·568-0·655) |
| **Subject-Level (Total)** | | | | | |
| COVID-19 vs. Non-COVID-19 (Max-Voting) | 0·943(0·920-0·965) | 0·953(0·899-1·000) | 0·836(0·799-0·873) | 0·851(0·818-0·884) | 0·632(0·547-0·717) |
| COVID-19 vs. Non-COVID-19 (Mean-Voting) | 0·938(0·916-0·961) | 0·781(0·677-0·885) | 0·918(0·891-0·944) | 0·900(0·872-0·927) | 0·676(0·585-0·767) |

AUC=Area Under the Curve. ACC= accuracy. Max=maximum. VS.=versus.

**Table 3. The confusion matrix of classification result of subjects and images on the test dataset of Shanghai, Hebei, Spain and the Total**

| | GT/Pred | Subject-level(max) | | | | Subject-level(mean) | | | | Image-level | | | |
|---|---|---|---|---|---|---|---|---|---|---|---|---|---|
| | | P | N | P(%) | N(%) | P | N | P(%) | N(%) | P | N | P(%) | N(%) |
| **Shanghai** | P | 29 | 0 | 100 | 0 | 25 | 4 | 86·2 | 13·8 | 99 | 40 | 71·2 | 28·8 |
| | N | 13 | 86 | 13·1 | 86·9 | 3 | 96 | 3 | 97 | 22 | 386 | 5·4 | 94·6 |
| **Hebei** | P | 26 | 1 | 96·3 | 3·7 | 21 | 6 | 77·8 | 22·2 | 100 | 35 | 74·1 | 25·9 |
| | N | 43 | 118 | 26·7 | 73·3 | 25 | 136 | 15·5 | 84·5 | 130 | 675 | 16·1 | 83·9 |
| **Spain** | P | 6 | 2 | 75 | 25 | 4 | 4 | 50 | 50 | 11 | 17 | 39·3 | 60·7 |
| | N | 12 | 142 | 7·8 | 92·2 | 6 | 148 | 3·9 | 96·1 | 23 | 601 | 3·7 | 96·3 |
| **Total** | P | 61 | 3 | 95·3 | 4·7 | 50 | 14 | 78·1 | 21·9 | 210 | 92 | 69·5 | 30·5 |
| | N | 68 | 346 | 16·4 | 83·6 | 34 | 380 | 8·2 | 91·8 | 175 | 1662 | 9·5 | 90·5 |